\documentclass[prl,twocolumn,showpacs,amsmath,amssymb,superscriptaddress]{revtex4-1}
\usepackage{graphicx}
\usepackage{hyperref}
\usepackage{breqn}
\usepackage{xcolor}
\usepackage{color}
\usepackage{booktabs}
\usepackage{float}
\usepackage{longtable}
\usepackage{epsfig}
\usepackage{bm}
\usepackage{dcolumn}
\usepackage{lipsum, babel}

\usepackage{rotating} % for sidewaystable

\usepackage[normalem]{ulem}
\usepackage{epstopdf}
\makeatletter
\let\cat@comma@active\@empty
\makeatother
\begin{document}

\title{Altermagnetic multiferroics and altermagnetoelectric effect}

\author{Libor \v{S}mejkal}

\affiliation{Max Planck Institute for the Physics of Complex Systems, N\"othnitzer Str. 38, 01187 Dresden, Germany}
\affiliation{Max Planck Institute for Chemical Physics of Solids, N\"othnitzer Str. 40, 01187 Dresden, Germany}
\affiliation{Institute of Physics, Czech Academy of Sciences, Cukrovarnick\'a 10, 162 00, Praha 6, Czech Republic}

\date{\today}

\begin{abstract}
Magnetoelectric multiferroics are highly sought after for applications in low-power electronics and for advancing fundamental research, including axion insulators and dark matter detection. However, achieving a combination of ferroic spin and electric orders, along with their controllable switching, remains a significant challenge in conventional ferromagnets and antiferromagnets. Here, we present first-principles evidence that time-reversal symmetry-breaking altermagnetic spin polarization with relatively high critical temperatures can emerge in ferroelectrics BaCuF$_4$ (T$_N$ $\sim$ 275K) and Ca$_3$Mn$_2$O$_7$ (T$_N$ $\sim$ 110K). Furthermore, we classify all possible altermagnetic polar spin groups, revealing altermagnetism in a collinear phase of BiFeO$_3$. We also propose an altermagnetoelectric effect, a nonrelativistic cross-coupling between altermagnetic spin polarization and ferroelectric polarization, mediated by a rotation of nonmagnetic polyhedra in the lattice structure. Our findings suggest an alternative pathway towards high-temperature magnetoelectric multiferroicity and the electric field control of altermagnetic order parameters.
\end{abstract}

\maketitle

% Word count without equations and figure captions: 2835
% figures: 5x aspsect ratio 1 gives: 5 x 170=850
% equations: 
% captions: 
% total: 
% PRL total limit 3750

%Paul

The magnetoelectric multiferroics, where electric fields induce magnetic order and vice versa, generated surge of renewed interest in past decades \cite{Spaldin2019, Fiebig2016, Khomskii2009} while remarkably its possible existence has been conjectured already in works by Pierre Curie\cite{Curie1894} and Landau and Lifshitz\cite{Landau1960}.
Ferroic orders describe spontaneously ordered phases, such as ferromagnetism or ferroelectricity, which can be reoriented by a conjugate magnetic or electric field. Multiferroics exhibit multiple ferroic properties simultaneously, while magnetoelectrics showcase coupling between magnetic and electric fields.

From a symmetry perspective, ferromagnetism occurs in materials with ferromagnetic point groups, while ferroelectricity is found in crystals with polar point groups. Although some polar point groups are compatible with ferromagnetism, the conditions required for the coexistence of ferroelectricity and ferromagnetism are often mutually contradictory \cite{Spaldin2019}.
Ferroelectric order typically arises from an electric dipole moment generated by Coulomb forces between displaced cations and anions within a polar crystal structure. However, as Coulomb forces are screened in metals, ferroelectricity prefers an insulating band structure and, consequently, fully filled electron shells. In contrast, ferromagnetism relies on unpaired electron spins, which is inherently favored in metallic, partially filled bands. This fundamental contraindication spurred renewed efforts since two decades ago to explore alternative pathways to achieve multiferroicity \cite{Spaldin2019,Fiebig2016,Khomskii2009}. 

Previously considered alternative mechanisms for achieving multiferroicity include composite multiferroics, single-phase multisublattice multiferroics - where one sublattice generates ferroelectricity and the other magnetism — and antiferromagnetic multiferroics where ferroelectricity is induced geometrically or by a noncollinear magnetism \cite{Spaldin2019,Fiebig2016,Khomskii2009}. While antiferromagnets are compatible with insulating electronic structure, antiferromagnetic order parameters lacking spin order in electronic structure couple only weakly to external fields\cite{Spaldin2019,Fiebig2016,Khomskii2009,Garcia2018}.

The family of magnetic materials has recently been expanded to include altermagnets by delineating altermagnets as an exclusively distinct spin-symmetry class of collinear magnets\cite{Smejkal2021a, Mazin2022a}. Altermagnets were revealed by a classificaiton based on spin group theory which employs pairs of (generally distinct) operations in spin and crystallogrpahic space\cite{Smejkal2021a}. 
Altermagnetism spontaneously breaks spin-rotational and half of the parent crystallographic symmetries while these crystallographic operations combined with spin-rotation are symmetries of altermagnets \cite{Smejkal2021a}. 
This unique spin group structure gives rise to a characteristic alternating and time-reversal symmetry breaking spin polarization in both direct (coordinate) and reciprocal (momentum) space\cite{Smejkal2021a, Smejkal2020, Mazin2021}, forming 
d-, g-, or i-wave patterns with 2, 4, or 6 spin-degenerate nodal surfaces \cite{Smejkal2021a, Smejkal2022a}.
These properties are rooted in the specific interplay of exchange interaction and crystal potential\cite{Smejkal2020} and demonstrate a ferroic character distinct from conventional ferromagnets and antiferromagnets\cite{Bhowal2024,McClarty2024,Fernandes2024}.
While numerous favorable characteristics of altermagnets have been already discussed in literature\cite{Smejkal2022a,Bai2024}, a conjugate field has not yet been identified that would strongly couple and thus enable an efficient control of the altermagnetic order. This motivates us to explore the interplay of altermagnetism with ferroelectricity.

In this Letter, we theoretically demonstrate the presence of altermagnetic spin polarization in ferroelectric polar materials and derive all possible altermagnetic polar spin groups. Additionally, we propose the altermagnetoelectric effect—a cross-coupling phenomenon between even-parity altermagnetic spin polarization and odd-parity ferroelectric polarization.
We show that the altermagnetoelectric coupling can microscopically originate from polyhedral rotations, illustrating that the same polyhedral distortions responsible for the so called geometric ferroelectricity can also generate altermagnetism. Using first-principles calculations, we confirm d-wave altermagnetic spin-splitting magnitudes of 25–200 meV in geometric ferroelectrics BaCuF$_4$\cite{Garcia2018} and Ca$_3$Mn$_2$O$_7$\cite{Benedek2011} across various electronic structure energy regions.
Our symmetry analysis identifies around two dozen candidate materials for altermagnetic multiferroics (see the full list in Supplementary information (SI)), including BiFeO$_3$, one of the most extensively studied multiferroics\cite{Spaldin2019,Haykal2020}.

 \begin{figure}[t]
	%\centering
%
	\includegraphics[width=0.5\textwidth]{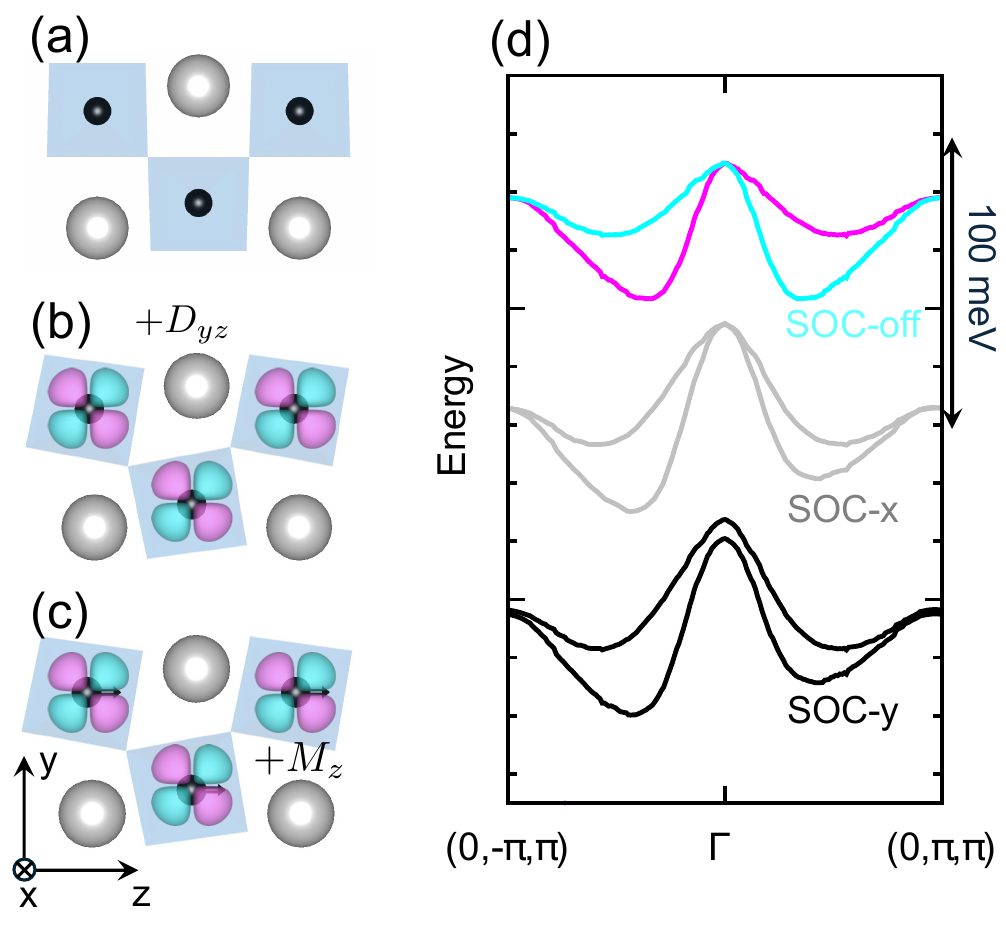} 
	\caption{\textbf{Ferroic altermagnetic order on polar crystal structure.} (a) Section of nonmagnetic high symmetric centrosymmetric crystal BaCuF$_4$ (Cu atoms marked in black, Ba grey, and F atoms occupie the corners of blue shaded octahedra). (b) Nonrelativistic ferroic altermagnetic $D_{yz}\sigma_{y}$-wave order generated by a geometric rotation of light-blue octahedra and antiferroic magnetic ordering on the Cu sublattices (see Fig.~3 for details). (c) Alermagnetism with additional relativistic spin-orbit coupling induced magnetization $M_{z}$. (d) Electronic band structure calculated without spin-orbit coupling, with spin-orbit coupling and moments along (001)-axis, and with spin-orbit coupling along (010)-axis.} 
	\label{fig1}
\end{figure}

\textit{Altermagnetism in ferroelectric polar crystal. --} We begin by demonstrating altermagnetic spin polarization and spin splitting in polar crystal structures that are either known or theoretically proposed to host ferroelectricity.
Fig.~1(a) shows the nonmagnetic, highly symmetric parent crystal structure of the geometric ferroelectric BaCuF$_4$ \cite{Fox1977,Garcia2018}. This structure exhibits a centrosymmetric point group, $mmm$. In contrast, Fig.~1(b) presents the polar crystal structure with the point group $mm2$, produced by rotations of the fluorine octahedra around the smaller Cu cation. These rotations also induce a ferroelectric shift of the Ba anions, allowing spontaneous electric polarization along the $z$-direction \cite{Garcia2018}.

Next, we examine the energy hierarchy of symmetry breaking in the electronic structure of BaCuF$_4$ with antiparallel (antiferromagnetic) ordering on the two Cu sublattices in the unit cell. We identify the spin point group symmetry of this magnetic state as $^{2}m ^{1}m^{2}2$. This symmetry indicates the presence of a $D_{yz}\sigma_{y}$, e.g.  $D_{yz}$-wave altermagnetic magnetization density that is ferroically ordered on the Cu sublattices with spin polarisation $\sigma_{y}$, as shown in Fig.~1(b,c). 
Remarkably, the altermagnetic density originates from the same geometric octahedral rotations responsible for ferroelectricity\cite{Garcia2018}. 
The generation of the altermagentic spin polarisation by rotated octahedra is similar to altermagentic mechanism in rutile structures\cite{Smejkal2020} and MnTe\cite{Smejkal2021a}. 
Our nonrelativistic calculations (see Methods and Ref.~\citep{Garcia2018}) confirm the presence of $D_{yz}\sigma_{y}$ altermagnetic order, as evidenced by the nonrelativistic altermagnetic spin splitting displayed in Fig.~1(d) (top, marked SOC-off). Around the $\Gamma$-point, this splitting exhibits a d-wave form,
$
k_{y}k_{z}\sigma_{y}
$.

When spin-orbit coupling (SOC) is included in our calculations, the spin point group symmetry is further reduced to a magnetic point group, which depends on the orientation of the Néel vector. For the Néel vector aligned along the $x$-axis, we obtain the magnetic point group $mm2$, which forbids a weak ferromagnetic moment. In stark contrast, when the Néel vector is aligned along the $y$-axis (the easy axis\cite{Garcia2018}), the magnetic point group becomes $m^{\prime}m^{\prime}2$, allowing an additional relativistic weak magnetization along the $z$-direction ($M_{z}$), as schematically illustrated in Fig.~1(c).
The effects of these symmetry breakings on the electronic band structure are captured in Fig.~1(d). Introducing SOC removes effective nonrelativistic mirror symmetry (due to spin-only group structure of collinear magnets\cite{Smejkal2021a}) in the band structure along two perpendicular directions in momentum space. This symmetry breaking results in weakly different electronic dispersions along these directions, as shown in Fig.~1(d) (middle, marked SOC-x). Furthermore, when the Néel vector aligns with the $y$-axis, the weak ferromagnetic moment introduces a ferromagnetic-like spin splitting at the $\Gamma$-point, confirmed in our band dispersion calculations in Fig.~1(d) (bottom, marked SOC-y).

\textit{Altermagnetoelectric effect. -- } The shared origin of altermagnetism and ferroelectricity in geometric motif rotations suggests a strong cross-coupling between these two orderings. This hypothesis is confirmed by our first-principles calculations, summarized in Fig.~2. A positive clockwise rotation of the fluorine octahedra (Fig.~2a) generates a positive ferroelectric polarization \cite{Garcia2018} as well as a positive altermagnetic spin polarization, as shown in Fig.~2b.
Reversing the sense of the octahedral rotation to counterclockwise is known to reverse the sign of the ferroelectric polarization\cite{Garcia2018}, as indicated in Fig.~2c. Our first-principles electronic structure calculations reveal that this reversal also flips the sign of the altermagnetic spin polarization, as shown in Fig.~2d. Thus, the signs of the ferroelectric polarization and the altermagnetic spin polarization are closely linked, with a change in the sign of one resulting in a corresponding change in the other. We term this cross-coupling of ferroic orders the \textit{altermagnetoelectric effect}.

Notably, unlike previously proposed relativistic couplings between weak ferromagnetic moments and ferroelectricity in BaCuF$_4$\cite{Garcia2018}, the altermagnetoelectric effect described here originates from nonrelativistic exchange interactions. Specifically, in this geometric ferroelectric, the altermagnetoelectric effect arises from the cross-coupling of nonrelativistic electric and magnetic-exchange fields within the unit cell, mediated by the rotated octahedral crystal motifs.

 \begin{figure}[t]
	\centering
	\includegraphics[width=0.490 \textwidth]{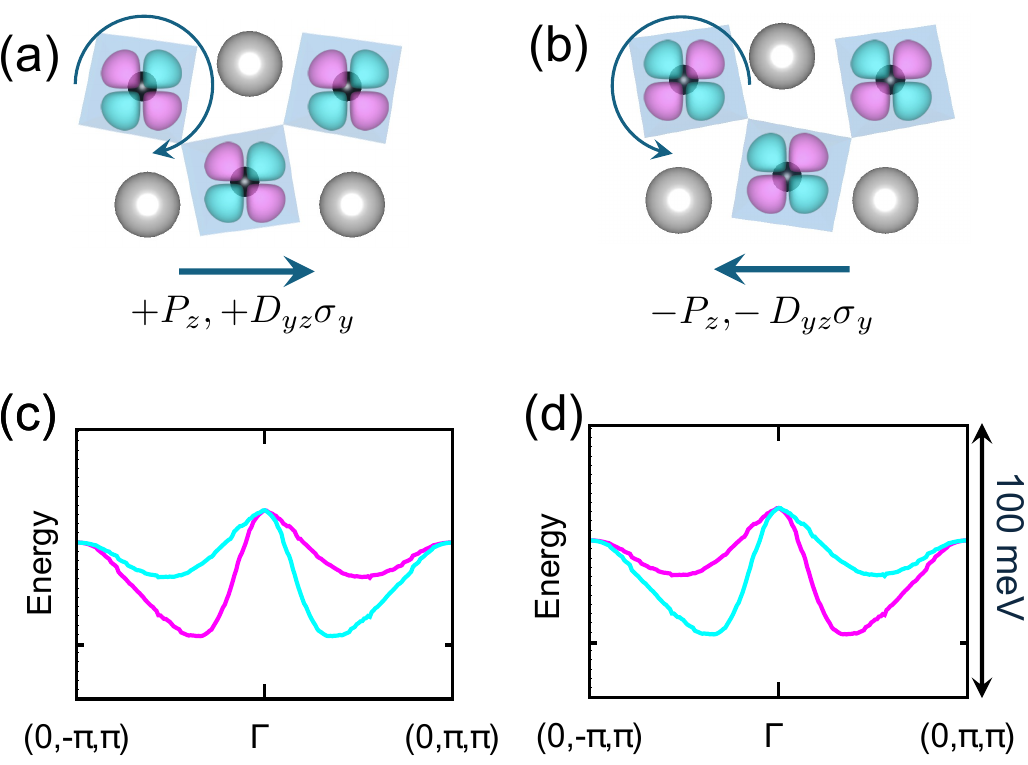} 
	\caption{\textbf{Altermagnetoelectric effect.} (a) Clockwise rotation of the blue-shaded octahedra generates a positive ferroelectric polarization and ferroic altermagnetic order on the Cu sublattices. (b) Density functional theory (DFT) calculated nonrelativistic altermagnetic spin splitting for the state described in (a). (c) Counterclockwise rotation of the blue-shaded octahedra reverses both the ferroelectric polarization and the ferroic altermagnetic order. (d) The reversal of the altermagnetic order is confirmed by the sign change of the corresponding spin splitting in the electronic structure calculated for the state in (c).} 
	\label{fig2}
\end{figure}

\begin{table}{}
\begin{tabular}{c|c|c}
\text {Class} & \text{Spin-point group} & \text{Material candidate}   \\
\hline\hline  \text { P-2 } & ${ }^2 m^1 m ^2 2$  &   \text{BaCuF$_4$}\cite{Garcia2018}   \\
\hline  \text { P-2 } & ${ }^2 m^2 m^1 2$  &   \text{Ca$_3$Mn$_2$O$_7$}\cite{Benedek2011}   \\
\hline  \text { P-2 } & ${ }^2 4$ &      \\
\hline  \text { P-2 } & ${ }^2 4^2 m^1 m$ & \text{BaMn$_2$V$_2$O$_8$}    \\
\hline  \text { B-2 } & ${ }^2 2$ &  \text{LiFeP$_2$O$_7$}  \\
\hline  \text { B-2 } &  ${ }^2 m$ & \text{Mn$_4$Nb$_2$O$_9$}  \\
\hline  \text { P-4 } & ${ }^1 4^2 m^2 m$ &  \text{Ba$_2$CoGe$_2$O$_7$}   \\
\hline  \text { B-4 } & ${ }^1 3 ^2 m$  &  \text{PbNiO$_3$}  \\ %\cite{Hao2013}
\hline  \text { B-4 } &  ${ }^2 6$   &  \\
\hline  \text { B-4 } &  ${ }^2 6^2 m^1 m$  & \text{Fe$_2$Mo$_3$O$_8$}\cite{Ghara2023}    \\
\hline  \text { P-6 } & ${ }^1 6^2 m^2 m$  & \text{BiFeO$_3$}\cite{Haykal2020}   \\
\end{tabular}
\caption{List of altermagnetic spin polarisation class (P - planar, B-bulk, for spin-momentum locking form see Ref.~\cite{Smejkal2021a}), polar altermagnetic spin point groups, and material candidates. The listed references contain the crystal and magnetic structures considered in the symmetry analysis. When no reference is provided the unit cell was taken from Magndata database \cite{Gallego2016}.}
\end{table}

\textit{Altermagnetic ferroelectric spin point group classification. -- } 
There exists 37 altermagnetic spin point groups which we enumerate in the SI\cite{Smejkal2021a}. Additionally, there are well-known 10 nonmagnetic polar point groups. Out of the 37 altermagnetic spin point groups, we identified 11 that are polar and we list them in the Tab~1.  Note that not all nonmagnetic polar point groups have counterparts in collinear altermagnetic polar spin point groups. For collinear altermagnetism cannot form from nonmagnetic polar point groups $1$ and $3$\cite{Smejkal2021a}. At the same time, there are some nonmagnetic polar point groups which allow for several possible altermagnetic polar spin point groups, e.g. $mm2$ can form two distinct altermagnetic types -   $^{2}m^{1}m^{2}2$, $^{2}m^{2}m^{1}2$. We will now discuss representative material candidates for these two spin-point groups. 

\textit{Altermagnetism in the proposed geometric ferroelectric BaCuF$_4$. --} We now present a more detailed discussion of the crystallographic, magnetic, and nonrelativistic electronic structure of BaCuF$_4$, calculated using first-principles methods. In Fig.~3(a), we depict the complete unit cell of ferroelectric BaCuF$_4$, highlighting the antiparallel alignment of magnetic dipoles on the Cu sites. The easy axis of magnetization is oriented along the $y$-axis.

Figure~3(b) illustrates the computed magnetization density around the two Cu atoms, Cu$_A$ and Cu$_B$, which possess opposite spins. The antiparallel alignment of spin dipoles on the Cu atoms gives rise to the spin  group symmetry $SO(2)\times Z_{2}\times^{2}m^{1}m^{2}2$\cite{Smejkal2021a} with nonntrivial spin point group elements:
\begin{equation}
\left[E\vert\vert E\right], \left[E\vert\vert M_{x}\right], \left[C_{2}\vert\vert M_{y}\right], \left[C_{2}\vert\vert C_{2z}\right].
\end{equation}
Here we use the same notation as introduced in Ref.~\cite{Smejkal2020}: $E$ marks identity in spin and crystal space, $C_{2}$ is a two-fold spin rotation and $M_{x}$, and $C_{2z}$ are mirror and two-fold rotation in crystallographic space. These spin symmetries stabilize the ferroic order of the d-wave altermagnetic magnetization component, $D_{yz}\sigma_{y}$ due to the spin symmetries $\left[C_{2}\vert\vert M_{y}\right]$, and $\left[C_{2}\vert\vert C_{2z}\right]$.

Finally, in Fig.~3(c), we plot the electronic band structure over a broader energy range. A significant spin-splitting magnitude, exceeding 120 meV, is observed in the energy region around $\sim$-1.5 eV. 
 \begin{figure}[t]
	\centering
	\includegraphics[width=0.490 \textwidth]{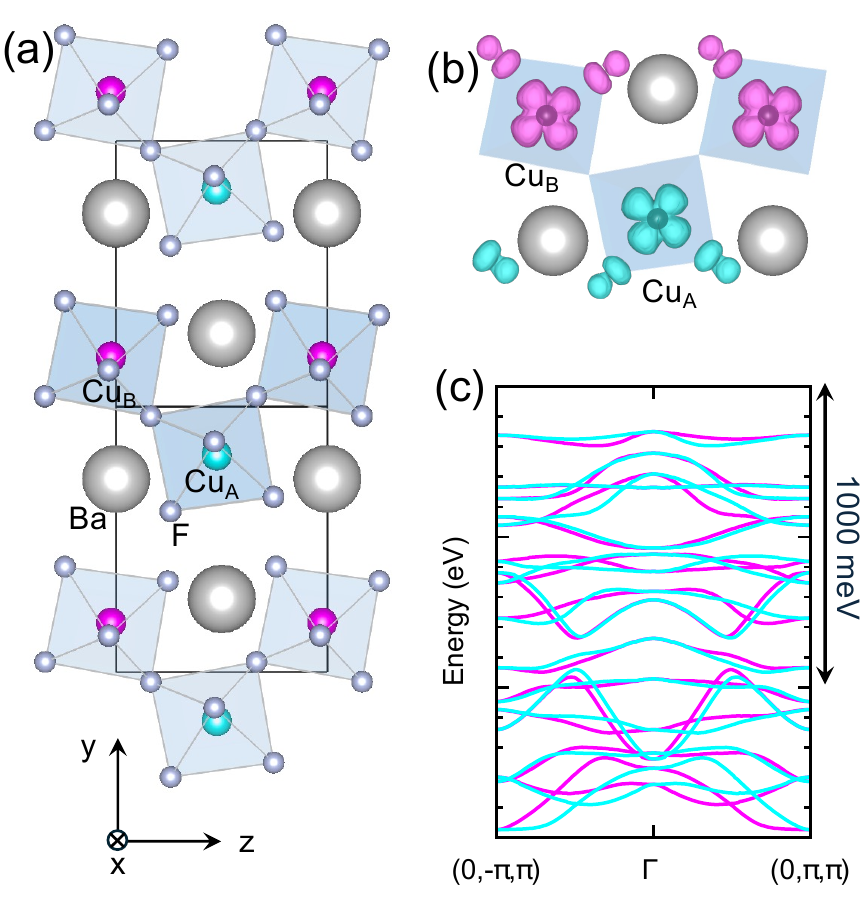} 
	\caption{\textbf{Altermagnetism in BaCuF$_4$.} (a) Crystal and antiferromagnetic structure. (b) First-principle calculated magnetisation density around Cu atoms. (c) Calculated nonrelativistic electronic bandstructure with altermagnetic spin splitting.} 
	\label{fig3}
\end{figure}

\textit{Altermagnetism in geometric ferroelectric Ca$_3$Mn$_2$O$_7$. --}
We have also identified altermagnetic electronic structure in Ca$_3$Mn$_2$O$_7$ crystal shown in Fig.~4(a). This material belongs to the Ruddlesden-Popper family of layered compounds, which are known to exhibit ferroelectric polarization with a magnitude on the order of $P_{S}$ $\sim$ 5 $\mu C/cm^{2}$, as predicted theoretically and confirmed experimentally\cite{Benedek2011,Oh2015,Liu2018}. The ferroelectricity in Ca$_3$Mn$_2$O$_7$ arises from a combined effect of octahedral tilt $X_{3}^{-}$ and rotation $X_{2}^{+}$, which are separately non-polar lattice modes \cite{Benedek2011}. Bellow  the Néel temperature $\sim$110 K, this material exhibits antiferromagnetic Néel vector $N_{z}$ oriented along z-direction on Mn sites marked by cyan and magenta colors in Fig.~4(a). 

In Fig.~4(b), we present our first-principle calculation (see Methods and Ref.~\cite{Rocha2020}) of the nonrelativistic anistropic magnetization density around the the Mn$_A$ and Mn$_B$ sublattices. We identify the spin point group of the magnetic state to be $SO(2)\times Z_{2}\times{ }^2 m^2 m^1 2$ with elements:
\begin{equation}
\left[E\vert\vert E\right], \left[C_{2}\vert\vert M_{x}\right], \left[C_{2}\vert\vert M_{y}\right], \left[E\vert\vert C_{2z}\right].
\end{equation}
This spin point group indicates a presence of a ferroically ordered d-wave altermagnetism $D_{xy}\sigma_{z}$ arising from the  anisotropic magnetization densities in real space. Our nonrelativistic band structure calculations, shown in Fig.~4(c) confirm the presence of corresponding $k_{x}k_{y}\sigma_{z}$ spin-splitting in the momentum space. The magnitude of the spin splitting can exceed 100 meV in the energy region around -2.5 to -2 eV below the Fermi level and when we switch-off Hubbard U in the calculations we observe splittings up to 200 meV. Finally, inclusion of spin-orbit coupling leads to magnetic point group $m^{\prime}m2^{\prime}$ and very weak ferromagnetic moment along x-axes $M_{x}$\cite{Benedek2011}. 

%Ca$_3$Mn$_2$O$_7$ : U = 4.5 eV, J=1 eV 4x4x2.

% 
 \begin{figure}[t]
	\centering
	\includegraphics[width=0.490 \textwidth]{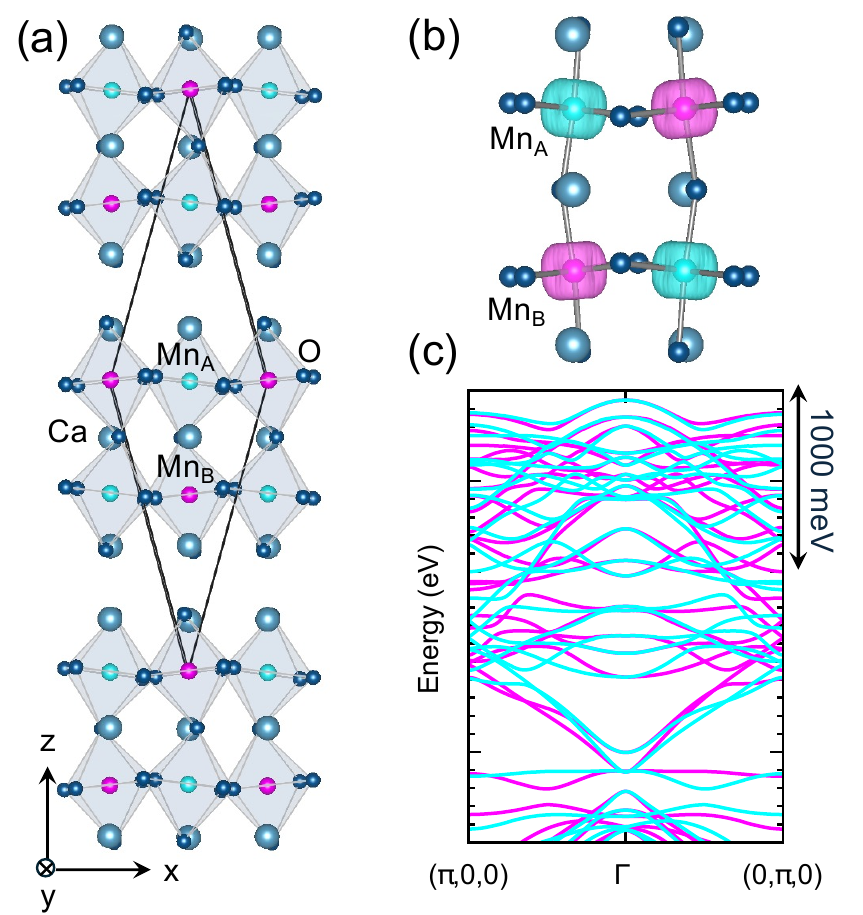} 
	\caption{\textbf{Altermagnetism in Ca$_3$Mn$_2$O$_7$.} (a) Crystal and antiferromagnetic structure. (b) First-principle calculated magnetisation density around Mn atoms. (c) Calculated electronic bandstructure with altermagnetic spin splitting.} 
	\label{fig2}
\end{figure}

\textit{Altermagnetic Magnetoelectric Multiferroic Materials --}
In Tab.~1, we observe that all altermagnetic polar groups are simultaneously also magnetoelectric spin point groups, which are characterized by the breaking of both parity/inversion and time-reversal symmetries. This is because altermagnetic spin densities inherently break time-reversal symmetry in addition to the inversion symmetry breaking necessary for a polar structure.
Tab.~1 further lists additional material candidates for altermagnetic magnetoelectric multiferroics. Beyond the previously discussed d-wave altermagnetic candidates, we also identify materials with g- and i-wave altermagnetic spin polarizations. Notably, g-wave altermagnetic materials such as Ba$_2$CoGe$_2$O$_7$ and Co$_2$Mo$_3$O$_8$ have already been associated with ferroelectricity \cite{Tokura2014, Ghara2023}. Interestingly, one of the most renowned multiferroic materials, BiFeO$_3$, is classified as an altermagnet in its collinear phase, according to our spin group symmetry analysis. This analysis also reveals i-wave altermagnetic spin polarization, which could explain the recently computed spin-polarised electronic strucutre of collinear BiFeO$_3$\cite{Bernardini2024}.
Although BiFeO$_3$ typically hosts an antiferromagnetic spin-spiral state, there are reports of a collinear phase under specific conditions, such as strain or the application of an electric field \cite{Haykal2020}.

\textit{Discussion. --} To summarize, we have theoretically demonstrated the altermagnetic spin polarisation in ferroelectrics and proposed the altermagnetoelectric effect. Our identification of ferroically ordered altermagnetism in ferroelectric materials with a high critical temperature, such as 275 K in BaCuF$_4$ and above room temperature in BiFeO$_3$, suggests a promising pathway toward room-temperature multiferroic magnetoelectrics—a potential solution to one of the grand challenges in the field of multiferroics\cite{Khomskii2009}. Furthermore, our theoretical prediction of the altermagnetoelectric effect opens the door to controlling altermagnetic ordering via an electric field.

The electric switching of ferroelectricity in Ca$_3$Mn$_2$O$_7$ has not yet been demonstrated experimentally. To address this issue, it was proposed that in BaCuF$_4$ (and similar systems\cite{Fox1977}), the weak ferromagnetic moment couples to the direction of ferroelectric polarization\cite{Garcia2018}. The presence of a weak ferromagnetic moment thus introduces an additional external control mechanism for studying the interplay between altermagnetism and ferroelectricity in this class of materials. A similar weak ferromagnetism–altermagnetism coupling was previously employed in MnTe altermagnet\cite{Smejkal2021a,Krempasky2024} to map and control altermagnetic domains\cite{Amin2024} and the anomalous Hall effect\cite{ Betancourt2021}.

An altermagnetic real-space order parameter and domains were recently mapped experimentally using X-ray magnetic circular dichroism, which exploits time-reversal symmetry breaking\cite{Smejkal2020} in MnTe altermagnetic semiconductors\cite{Amin2024}. Experimental confirmation of electric field control over altermagnetism could be achieved using similar techniques. Coupling altermagnetic order to external fields remains one of the primary challenges in the emerging field of altermagnetism and altermagnetic spintronics. Our proposed altermagnetoelectric effect offers a sought-after mechanism for the electric control of altermagnetism, paving the way for novel functionalities in altermagnetic materials.

\textit{Acknowledgement. --}
We acknowledge funding by the Deutsche Forschungsgemeinschaft (DFG, German Research Foundation) TRR 288 – 422213477 (project A09 and B05), and support by the Ministry of Education of the Czech Republic CZ.02.01.01/00/22008/0004594.

%\bibliography{export}


\end{document}